# A two-dimensional terahertz smart wristband for integrated sensing and communication

Shaojing Liu, Yongsheng Zhu, Runli Li, Ximiao Wang, Hongjia Zhu, Shangdong Li, Hai Ou, Yanlin Ke, Runze Zhan, and Huanjun Chen[*]

State Key Laboratory of Optoelectronic Materials and Technologies Guangdong Province Key Laboratory of Display Material and Technology, School of Electronics and Information Technology, Sun Yat-sen University, Guangzhou, China
E-mail: chenhj8@mail.sysu.edu.cn

The development of wearable devices for terahertz (THz) integrated sensing and communication (ISAC) is pivotal for forthcoming 6G Internet of Things (IoT) and wearable optoelectronics. However, existing THz system suffers from bulkiness, narrow spectral response and limited flexibility constrained by their dependence on external antennas, complex coupling architectures and rigid components. Here, we present a 2D THz smart wristband based on a graphene plasmon polariton atomic cavity (PPAC) array, which integrates sensing and communication within a monolithic microdetector. Operating without any external antenna, the compact and flexible device enables self-powered, polarization-sensitive and frequency-selective THz detection across a broad response spectrum from 0.25 to 4.24 THz, with a responsivity of 6 V/W, a response time of 62 ms, and mechanical robustness maintained over 2000 bending cycles. Notably, we further exploit its multi-parameter THz responses for dual-purpose ISAC functionality. For sensing, the polarization- and strain-dependent THz responses are utilized as high-dimensional features for a convolutional neural network (CNN), enabling circuit fault diagnosis with 97% accuracy. For communication, the device implements secure encrypted communication under simulated on-body wearing condition through dual-channel encoding of THz polarization and on-off signals. This 2D ISAC platform paves the way for miniaturized, intelligent wearable systems for advanced human–machine interaction.

Terahertz (THz) waves (0.1–10 THz), characterized by abundant spectral resources, strong penetration and wide bandwidth, are anticipated to become a key technology for sixth-generation (6G) mobile networks[1–3]. This will propel the evolution of the Internet of Things (IoT) from a conventional communication system toward an integrated intelligent platform capable of simultaneous communication, sensing and computing. Wearable devices with integrated sensing and communication (ISAC) functionalities in THz regions are particularly promising due to their multifunctionality and flexibility, enabling adapting to complex contours for efficient environmental perception, target identification and high-speed data transmission across diverse scenarios[4–6]. For instance, wearable devices or "electronic skin" worn on the curved surfaces or joints of robots could detect concealed weapons or assess chip quality, while lightweight flexible sensors integrated onto objects with arbitrary geometries enable high-speed wireless communication and continuous monitoring. However, existing THz detection technologies still face a significant gap in meeting the pressing demands for flexibility, miniaturization, broadband response and multifunctional integration in wearable devices[7–10]. First, conventional THz detectors rely on rigid and brittle semiconductor materials like high electron mobility transistors (HEMTs) and Schottky barrier diodes (SBDs), which are incompatible with flexible and deformable substrates required for wearables. Second, mainstream THz detectors available on the market or under research—including Golay cells, thermopiles, and bolometers—often require external antennas, complex coupling structures or cryogenic cooling system to enhance performance, leading to large device footprints, limited detection parameters and intricate fabrication processes. Therefore, developing a flexible THz micro/nano detection system capable of multiparameter-sensitive sensing and communication functionalities remains a critical challenge.

Van der Waals two-dimensional (2D) materials, represented by graphene, provide an ideal platform for constructing next-generation flexible and miniaturized THz detectors due to their atomic thickness, superior mechanical flexibility, and tunable optoelectronic properties[11–13]. Recent advances in flexible THz detectors based on 2D materials, such as graphene, WTe$_2$, or T$_d$-MoTe$_2$, have been demonstrated remarkable responsivity, detectivity and response speed[14–16]. Nevertheless, owing to the weak light-matter interaction between long-wavelength THz radiation and ultrathin materials, most devices often incorporate metallic antennas, metasurfaces or large channel length to enhance anisotropic THz absorption and the hot-carrier generation efficiency. This antenna-coupled architecture not only enlarges the device footprint, but also fundamentally limits the response frequency below 1 THz, preventing the detection of high-frequency THz signals crucial for future 6G communications. Such challenges, including narrow spectral response, limited polarization sensitivity and bulky system, are contrary to the multifunctionality, miniaturization and flexible desired for wearable applications. A monolithic strategy—free of additional optical components—that enables both broadband response and polarization selectivity in THz region is therefore highly desirable.

Graphene plasmons, collective excitations of 2D massless electrons, enable extensive field confinement and enhancement in THz regions without traditional antennas and metasurfaces[17–19]. These plasmons decay inelastically within a few femtoseconds to generate hot carriers, which in turn undergo rapid electron-electron energy transfer to convert into measurable electrical signals. Patterning monolayer graphene into periodic micro/nanostructures allows direct far-field excitation of graphene plasmon polaritons resonance under THz irradiation, with resonance intensity, frequency and polarization tunable via geometry, doping, and dielectric environment[20–22]. Efficient converting graphene plasmons and their decay into discernible THz response is crucial for compact, room-temperature and multi-parameter detection. Building on this, we recently proposed a novel THz detection mechanism based on graphene *plasmon polariton atomic cavities* (PPACs), enabling highly sensitive, polarization-sensitive and spectrally selective THz detection on rigid substrate without external antennas[23]. However, integrating graphene plasmon polaritons into wearable ISAC systems still faces multiple challenges, including flexible adaptation and performance stability under mechanical strain.

In this study, we report a flexible and compact 2D THz smart wristband designed for ISAC applications. Its core microdetector, fabricated on a flexible substrate, features an asymmetric graphene channel comprising graphene rectangle microarray—acting as PPACs—and unpatterned graphene. The rectangle PPACs are designed to support plasmonic resonance with exceptional selectivity in intensity, frequency and polarization across the 0.25–4.24 THz range, with resonance behavior tunable via the dimensions and applied mechanical strains. Notably, the flexible ISAC microdetector retains stable mechanical flexibility and THz response performance even after 2000 bending cycles. Leveraging these properties, we demonstrate, for the first time, a monolithic 2D flexible microdetector that integrates both THz sensing and communication. To showcase its sensing capability, we extract THz response signals of the device under varying polarization angles and strains as features to train a convolutional neural network (CNN), achieving 97% accuracy in non-destructive circuit fault classification. In addition, for communication, we develop a dual-channel encrypted transmission system by exploiting the dual sensitivity of the device to THz polarization angles and on-off states, thereby effectively securing critical information transmission for the IoT. This 2D flexible microdetector integrating THz sensing and communication functionalities offers novel strategies for device design and system integration, paving the way for next-generation intelligent, flexible, and miniaturized ISAC terminals tailored for the 6G IoT era.

## Sensing and communication functionalities of 2D THz ISAC smart wristband

Figure 1a illustrates the 2D THz ISAC smart wristband worn on a curved human wrist (top panel), alongside a schematic of its core microdetector (bottom panel) and a corresponding photograph (Supplementary Fig. 1). This flexible ISAC microdetector features a two-terminal configuration in which the source and drain electrodes are bridged by an asymmetric graphene channel. The graphene channel comprises a periodic array of graphene rectangles—functioning as PPACs—integrated with an unpatterned graphene region. Upon THz illumination, the THz response is collected across the source–drain terminals at zero bias. Attributed to its unique structural design, the device exhibits highly sensitive responses to THz intensity, frequency, polarization, and to mechanical strain induced by substrate bending. Leveraging the multi-parameter sensitivity of the device, we achieve the monolithic integration of THz sensing and communication functionalities on a 2D flexible ISAC microdetecor. For sensing and identification, THz waves transmitted through defective regions of the internal circuit in the chip are detected under varying polarization angles and mechanical strains. The resulting signals are processed by a CNN to achieve accurate fault classification and identification (Fig. 1b). For secure communication, a dual-channel encrypted communication system is implemented by exploiting the dual sensitivity of the device to THz polarization angle and on-off states (Fig. 1c). The device receives a superimposed encrypted signal in which one channel carries the raw message and the other delivers the security key. Decryption is subsequently performed by processing the received encrypted signal with the extracted key to recover the raw message. This design demonstrates the potential of integrating intelligent sensing and secure communication on a unified flexible platform, offering a novel paradigm for future wearable integrated systems

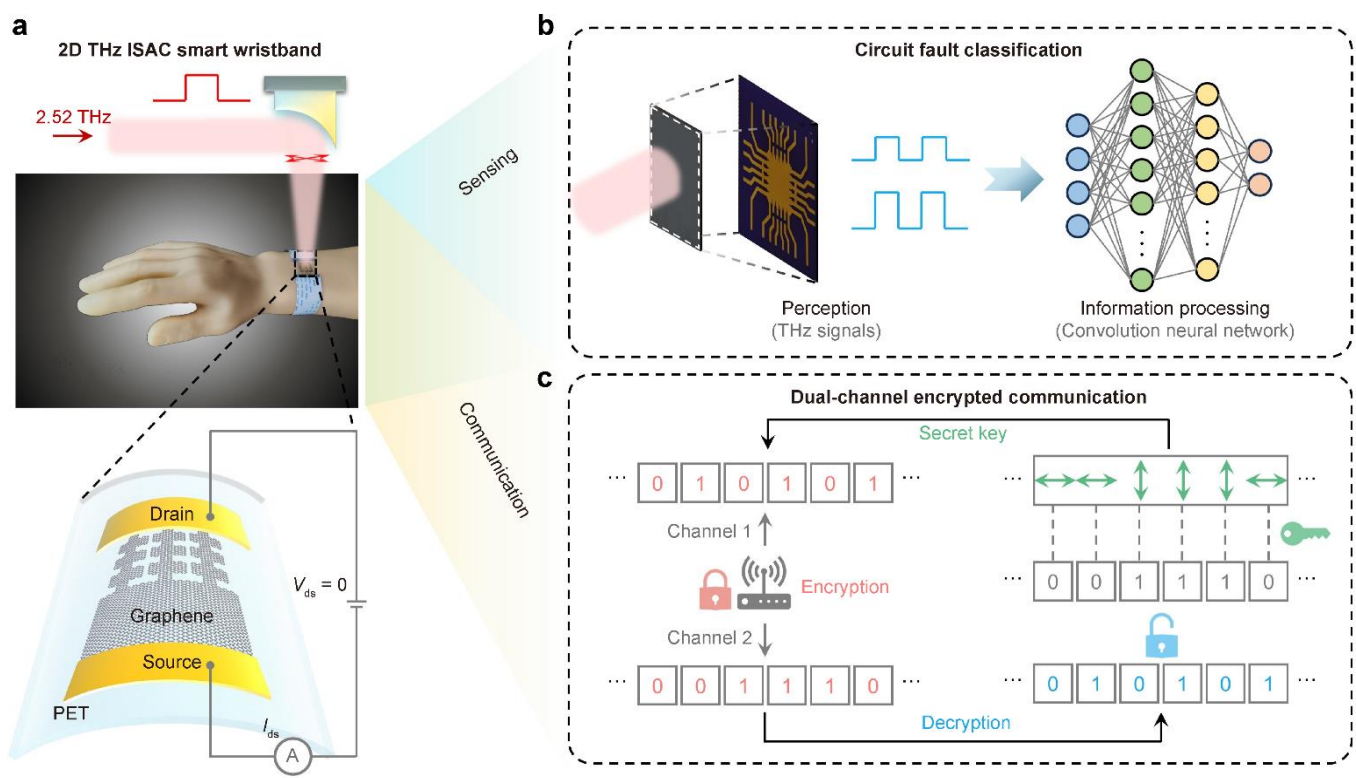

**Fig. 1 | Sensing and communication functionalities of 2D THz ISAC smart wristband. a**, 2D THz ISAC wristband, consisting of an asymmetric graphene channel connected by source and drain electrodes, is worn on a simulated curved surface of the human wrist. The THz signal can be obtained from source–drain current ($I_{ds}$) under zero bias, which serves for the input of CNN and the reception of dual-channel encrypted communication, enabling integrated sensing and communication. **b**, Circuit fault classification by the 2D THz ISAC wristband. The THz signals carrying information about internal circuit fault in the chip are received by the 2D THz ISAC wristband and fed into a CNN for training, enabling the identification and classification of circuit faults. **c**, Dual-channel encrypted communication by the 2D THz ISAC wristband. The wristband receives encrypted THz signals superimposed with the raw message and a cryptographic key.

## Design, fabrication and characterization of flexible graphene rectangle PPAC

In the proposed 2D THz ISAC smart wristband, an asymmetric graphene channel composed of rectangle PPACs and unpatterned graphene is used as the THz photosensitive layer. The electromagnetic field confinement and absorption within the PPACs are significantly enhanced when the incident THz frequency matches the resonance frequency of plasmon polariton resonance in the PPACs. The absorption intensity, resonance frequency and polarization characteristics of the plasmonic response can be tuned by varying the aspect ratio (AR) and applied strain ($\varepsilon$) of the PPACs. To systematically investigate the influence of these parameters on the plasmon polariton resonance behavior, we employed finite-difference time-domain (FDTD) simulations and THz time-domain spectroscopy (THz-TDS). In both the simulation and experiment, THz waves were perpendicularly incident on the PPACs with the polarization angle ($\theta$) set to 0° (*x*-polarization), aligning the electric field with the long axis of the PPACs (Fig. 2a).

Figures 2c and 2d present the simulated absorption spectra and the dependence of the resonance frequency on the aspect ratio (AR = *L* / *W*). Each PPACs exhibits a distinct plasmonic resonance peak, which redshifts across the 0.1–5 THz range as the AR increases from 0.7 to 30. Simulated near-field distributions reveal strong electric field localization along the long axis of the rectangle PPACs, as depicted in the inset of Fig. 1d. Furthermore, the dependences of absorption spectra and resonance intensity on the polarization angle are investigated. As demonstrated in Fig. 1e, while the resonance frequency associated with the long axis of the PPACs remains largely invariant with varying $\theta$, the corresponding resonance intensity attenuates as $\theta$ increases from 0° to 90°. This behavior is characteristic of dipolar plasmonic resonance, where excitation efficiency is maximized when the incident electric field aligns with the long axis of the PPACs. The polar plot of resonance intensity further elucidates this dipolar behavior (Fig. 1f)—the resonance intensity peak appears at $\theta$ = 0° and diminishes as $\theta$ increases, falling to nearly zero as $\theta$ approaches 90°. At $\theta$ = 90°, the electric field component along the long axis of the PPACs vanishes, effectively suppressing plasmon excitation. This polarization-dependent behavior highlights a key advantage of the rectangle PPAC design. In contrast to conventional antenna-coupled graphene micro/nanostructures, which typically require complex, bulky antenna designs to support multi-polarization or frequency-dependent responses, the simple rectangle geometry of the PPACs provides comparable functionality. With a compact and monolithic architecture, the PPAC enables tunable and polarization-sensitive responses across a broad frequency range, significantly reducing both design complexity and device footprint.

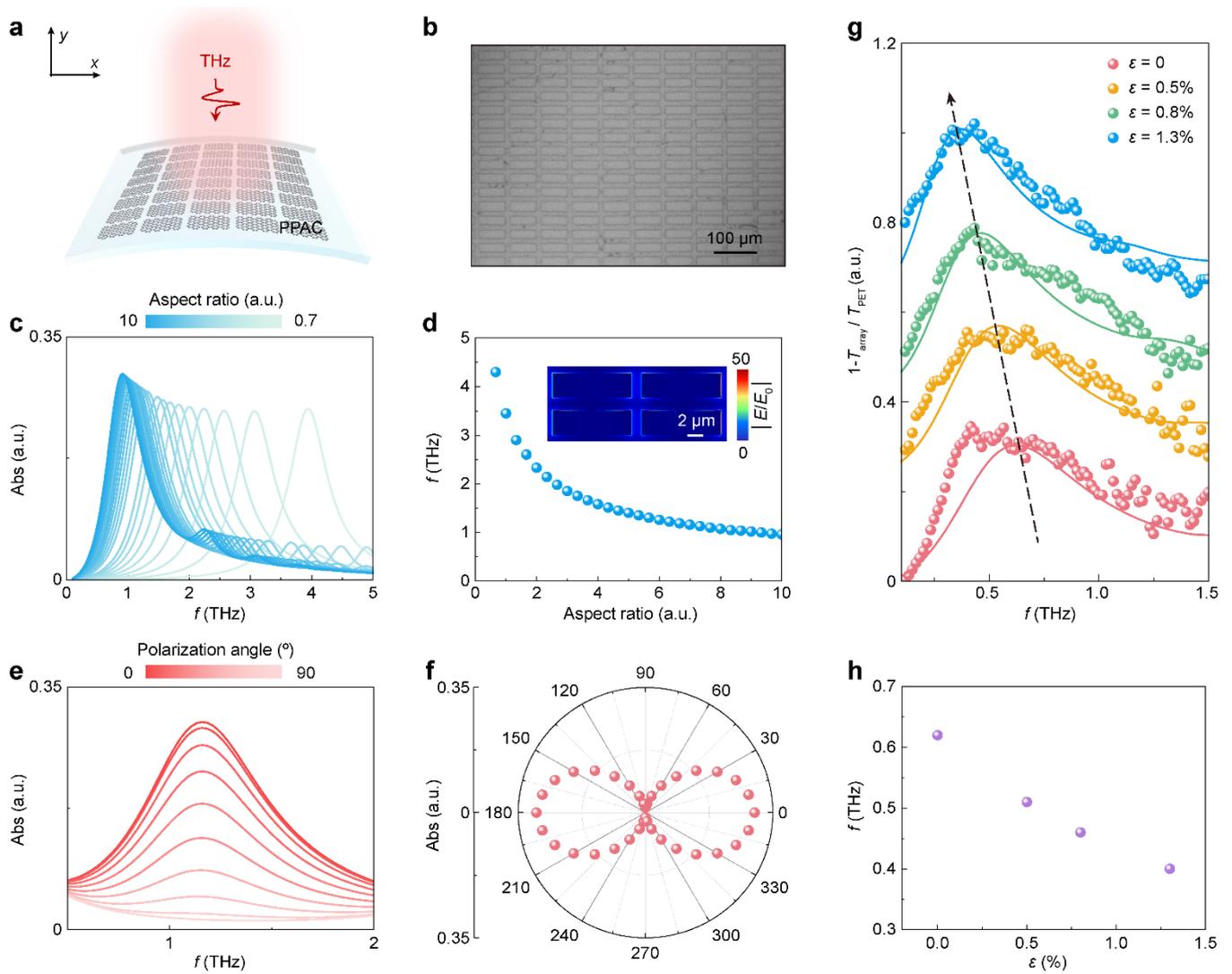

**Fig. 2 | Flexible graphene rectangle PPACs and their tunable plasmonic resonance characteristic. a**, Schematic illustration of the graphene rectangle PPAC array on the flexible PET substrate. **b**, Optical microscopy image of rectangle PPAC array with dimensions of 60 μm × 20 μm. **c**, Simulated THz absorption spectra of the rectangle PPAC array with their aspect ratio ranging from 0.7 to 30. **d**, Simulated resonance frequency (*f*) of the rectangle PPAC array as a function of the aspect ratio (AR). The inset shows simulated electric field distribution of the rectangle PPAC array with dimensions of 10 μm × 3 μm. **e**, Simulated THz absorption spectra of the rectangle PPAC array at different excitation polarization angles. The polarization angle is defined as the angle between the polarization direction of the incident THz wave and the long axis of the rectangle PPAC. **f**, Simulated polarization polar plot of the absorption intensity for PPAC array against the polarization angle, which displays a dipolar plasmonic resonance behavior. **g**, Experimental (dots) and simulated (lines) THz absorption spectra of the rectangle PPAC array in **b** with varying uniaxial strain (*ε*) applied along the long axis of the rectangle PPAC, showing the resonance frequency shift with changing strain. **h**, Experimental resonance frequencies as a function of the strain, extracted from **g**.

In addition to geometric aspect ratios, the resonance frequencies of graphene PPACs fabricated on flexible substrates, unlike on rigid substrates, can be dynamically tuned via mechanical strain. The applied strain ($\varepsilon$) is defined as[24],

$$\varepsilon = \frac{h}{2r} \times 100\% \tag{1}$$

where $h$ denotes the substrate thickness and $r$ represents the bending radius of curvature. Mechanical strain alters both the geometric aspect ratio and the Fermi level of the rectangle PPACs, thereby affecting their plasmonic resonance characteristics. Specifically, an increase in uniaxial strain applied parallel to the long axis of rectangle PPACs induces a progressive redshift in the resonance frequency upon the irradiation of $x$-polarized THz waves. This behavior occurs because the increasing transverse strain simultaneously enlarges the aspect ratio and reduces the Fermi level, resulting in a lower plasmon oscillation frequency and a consequent redshift of the resonance peak. To experimentally validate these findings, the absorption spectra of rectangle graphene PPACs under varying transverse strains were characterized using a THz-TDS. Raman measurements were first conducted to evaluate the quality and uniformity of the monolayer graphene. Due to the intrinsic Raman signal from polyethylene terephthalate (PET) substrate, the monolayer graphene grown by chemical vapor deposition (CVD) was transferred onto a Si/SiO$_2$ substrate for characterization by confocal Raman spectroscopy (Supplementary Fig. 2a). The resulting Raman spectrum exhibits two distinct characteristic peaks, which are the 2D peak at ~2700 cm$^{-1}$ and the G peak at ~1580 cm$^{-1}$, with an intensity ratio $I_{2D}/I_G$ approximately equal to 1.84. These results confirm the high crystallinity and monolayer nature of the graphene. We also performed Raman measurements on the graphene transferred onto the PET substrate. As depicted in Supplementary Fig. 2b, subtracting the intrinsic Raman spectrum of the PET substrate (blue curve) from that of graphene on the PET substrate (red curve) clearly reveals the characteristic 2D and G peaks of monolayer graphene at ~2700 cm$^{-1}$ and ~1580 cm$^{-1}$, respectively. Subsequent Raman characterization of the rectangle PPAC array under varying transverse strains indicates a redshift in the 2D peak with increasing transverse strain (Supplementary Fig. 3). This shift arises from the elongation of carbon–carbon bonds, which reduces bond energy and consequently lowers the lattice vibration frequency.

To ensure sufficient strong absorption signals, PPAC array with unit dimensions of 60 μm × 20 μm were fabricated on monolayer graphene transferred onto a PET substrate (see Methods), as shown in the optical microscope image in Fig. 1d. We measured the THz absorption spectra of the rectangle PPAC array under different transverse strains, with the incident polarization aligned parallel to the long axis of the PPACs. The absorption spectrum can be obtained as,

$$Abs = 1 - \frac{T_{array}}{T_{PET}} \tag{2}$$

where $T_{PPAC}$ is the transmission of the graphene PPAC array on the PET substrate, and $T_{PET}$ is the transmission of the blank PET substrate. The resulting spectra, presented in Fig. 1g, obviously demonstrate resonance features consistent with the simulation predictions—the resonance frequency of the graphene rectangle PPAC arrays undergoes a redshift as transverse strain increases (Fig. 1h). Overall, compared with high-symmetry geometries such as squares, circles, or hexagons fabricated on rigid substrates, the flexible rectangle PPACs offer greater tunability due to both tunable aspect ratios and mechanical strain. This capability enables a more flexible and broader tuning range for both resonance frequency and polarization sensitivity, which introduces a new degree of freedom to the design of graphene plasmonic devices.

## Fabrication and characterization of 2D THz ISAC smart wristband

By coupling plasmon polariton resonance with the photothermoelectric (PTE) effect in rectangle PPACs, sensitive and zero-bias detection of incident THz waves with specific frequencies and polarization states can be achieve[23]. To this end, we developed a 2D THz ISAC smart wristband featuring a flexible ISAC microdetector as its core component. The device utilizes a two-terminal configuration in which the source and drain electrodes are bridged by an asymmetric graphene channel, as shown in Fig. 3a. The graphene channel was fabricated using monolayer graphene transferred onto a PET substrate and subsequently patterned via ultraviolet (UV) maskless lithography. The channel comprises a rectangle PPAC array interconnected by graphene microribbons on the left side—with individual PPAC dimensions of 30 μm × 10 μm—and an unpatterned graphene region with the size of 65 μm × 100 μm on the right side, yielding a total channel length ($L_c$) of approximately 140 μm. The source and drain electrodes were defined using aligned UV maskless lithography followed by Cr/Au (10 nm/100 nm) deposition and a lift-off process. The absorption spectrum of the PPAC array exhibits a pronounced resonance peak at 1 THz within the 0.1–5 THz range (Supplementary Fig. 4a), facilitating efficient coupling with incident THz radiation. Under dark conditions, the current–voltage (I–V) characteristics display a linear relationship, confirming good Ohmic contact between the graphene and the electrodes, which is essential for high-performance THz detection (Supplementary Fig. 4b).

To validate the THz response mechanism of the flexible ISAC microdetector under resonance excitation, the electric field of the device was investigated using FDTD simulations (Supplementary Fig. 5). Under THz illumination, the THz response of the device arises from two coupled physical mechanisms—(i) PPAC-mediated field enhancement and hot-carrier generation, and (ii) the generation of a zero-bias photovoltage driven by asymmetric carrier/temperature gradients. Specifically, the plasmonic resonance modes within individual PPAC unit strongly confine the incident THz field at deep-subwavelength scales, significantly enhancing THz absorption in the graphene and thus generating a high density of hot carriers. Subsequently, the asymmetric absorption profile across the graphene channel establishes a hot-carrier concentration gradient, resulting in a non-uniform temperature distribution $T(x)$ along the channel. Driven by these gradients, under uniform illumination, the PTE effect drives the diffusion of hot carriers from the PPAC array region toward the unpatterned graphene region, where they are ultimately collected by the electrodes. This process generates a net zero-bias photovoltage $V_{ph}$ across the device terminals, which is calculated by integrating the potential gradient $\nabla V(x)$ along the channel length $L_c$[25],

$$V_{ph} = -\int_0^{L_c} S(x)\nabla T(x)dx \qquad (3)$$

$$S(x) = -\frac{\pi^2 k_B^2 T(x)}{3e}\left(\frac{d\ln\sigma}{dE}\right)\bigg|_{E=E_f} \qquad (4)$$

where $L_c$ is the channel length, $S(x)$ is the Seebeck coefficient, $\nabla T(x)$ is the temperature gradient, $k_B$ is Boltzmann constant, $T(x)$ is absolute temperature, $e$ is the elementary charge and $\sigma$ is electrical conductivity of graphene. This theoretical model elucidates the microscopic mechanism governing photon-to-electron conversion within the device.

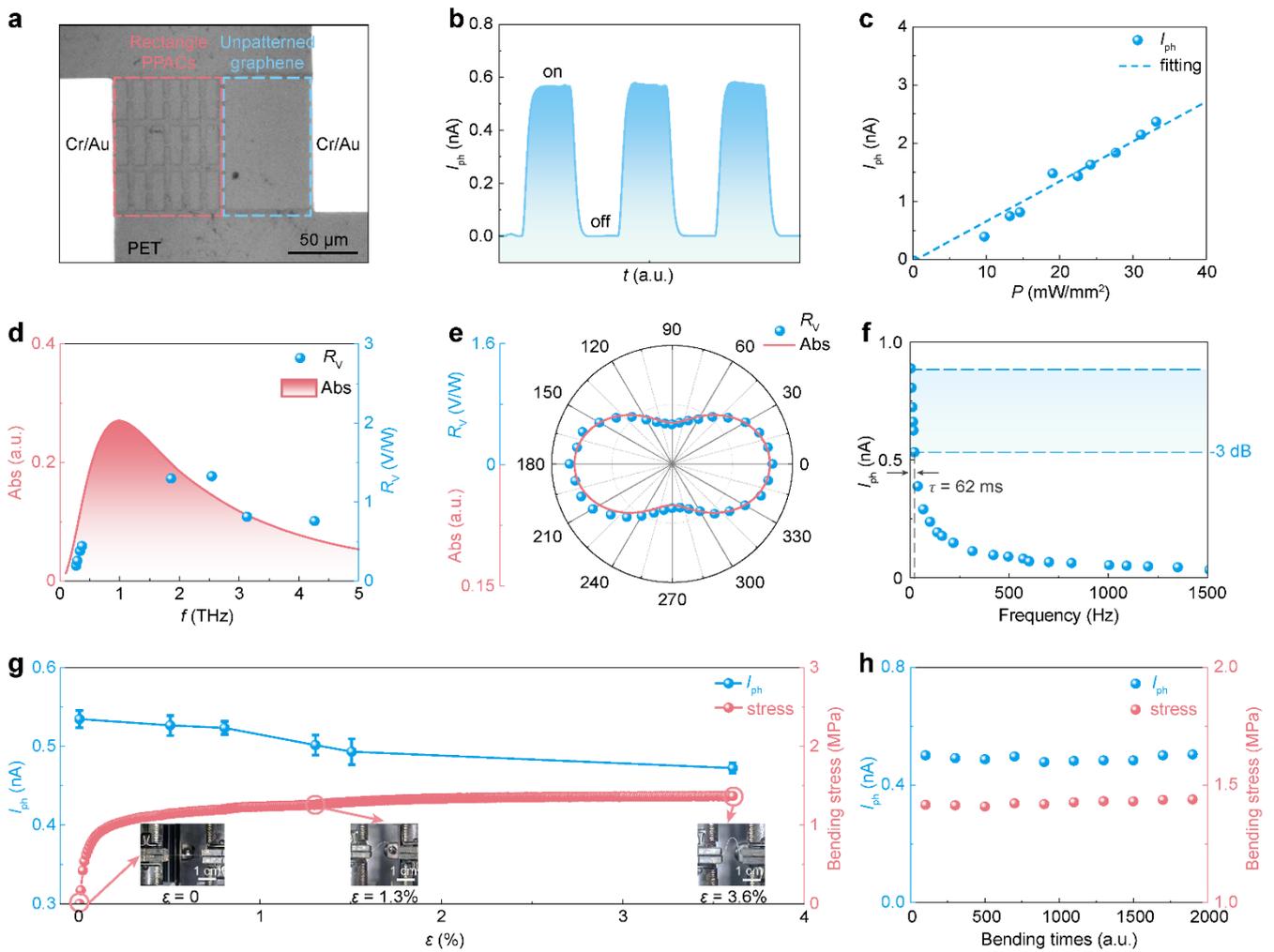

**Fig. 3 | Fabrication and characterization of flexible ISAC microdetector. a**, Optical microscopy of the flexible ISAC microdetector, consisting of two Cr/Au metal electrodes connected by an asymmetric graphene channel fabricated on the PET substrate. The asymmetric channel comprises a graphene rectangle PPAC array connected by graphene microribbons on one side and unpatterned graphene on the other. **b**, Photocurrent of the device under periodic 2.52 THz illumination. **c**, Photocurrent as a function of illumination power. The dots and dash line illustrate the experimental data and linear fits, respectively. **d**, Frequency-selective photoresponse (dots) and

simulated THz absorption spectrum (solid line) of the device. **e**, Polarization-resolved photoresponse (dots) and simulated polarization polar plot (solid line) of the device. **f**, Variation of the photocurrent with modulation frequency. The reciprocal of the modulation frequency at which the photocurrent falls to -3dB is defined as the response time. **g**, The bending stress and photocurrent of the device with different uniaxial bending strain applied on the device under 2.52 THz illumination. The insets present the photographs of the device at $\varepsilon = 0$, $\varepsilon = 1.3\%$ and $\varepsilon = 3.6\%$ from left to right panel, respectively. **h**, The bending stress and photocurrent of the device during 2000 bending times under 2.52 THz illumination with the bending strain of 1.3%.

To evaluate the THz response, a high-power 2.52 THz laser was employed as the excitation source. The THz response of the flexible ISAC microdetector was characterized under 2.52 THz radiation linearly polarized parallel to the long axis of the rectangle PPACs. As depicted in Fig. 3b, a pronounced and highly reproducible photocurrent signal was generated at zero bias. Power-dependent measurements reveal a linear relationship between the photocurrent and incident power (Fig. 3c). The voltage responsivity ($R_V$) can be expressed by,

$$R_V = \frac{V_{ph}}{P_{eff}} = \frac{I_{ph} \times R}{P_{eff}} \tag{5}$$

$$I_{ph} = \frac{2\pi\sqrt{2}V_{lock}}{4G} \tag{6}$$

where $V_{ph}$ and $I_{ph}$ denote the photovoltage and photocurrent at a specific illumination power, respectively. $R$ is the resistance of the graphene channel, $V_{lock}$ is the voltage readout from the lock-in amplifier, $G$ is the preamplifier gain in V/A, and $P_{eff}$ is the effective power. Notably, the beam diameter of the THz source (~1 mm) is significantly larger than the effective channel area of our antenna-free detector. This stands in stark contrast to conventional antenna-integrated THz detectors, where the effective device area is typically comparable to the beam size. Consequently, $P_{eff}$ is determined by the total incident power, the laser spot size and the channel area, defined as,

$$P_{eff} = \frac{P_0 \times S_{device}}{S_0} \tag{7}$$

where $S_{device}$ represents the effective area of the device (*i.e.*, the graphene channel area), while $P_0$ and $S_0$ denote the incident THz power and beam spot size, respectively. For a channel length of 140 μm, the calculated $R_V$ reaches 1.3 V/W at 2.52 THz, approximately a 40-fold enhancement compared to our previously reported flexible photodetectors at the same frequency[26]. This significant improvement is attributed to the plasmonic resonance of the rectangle PPACs, which facilitates efficient electric field confinement and localized absorption of incident THz radiation within the graphene channel. Furthermore, our previous research established that scaling down channel lengths effectively boosts responsivity[26]. To further optimize performance, we reduced the channel length from 140 μm to 20 μm while maintaining a constant channel width (Supplementary Fig. 6a). The resistance, derived from I–V characteristics, decreases linearly from 333 to 291 kΩ as the channel lengths drop from 140 μm to 20 μm (Supplementary Fig. 6b), underscoring the superior conductivity of the graphene channel. Under 2.52 THz illumination, the 20-μm device achieved an $R_V$ of 6.5 V/W, a remarkable value for flexible THz detectors based on low-dimensional materials (Supplementary Table 1)[14–16, 26–33]. While lots of reported flexible 2D THz detectors exhibit high responsivity, they typically rely on metallic antennas or metasurfaces to induce electric field localization and enhancement. These additional structures inevitably enlarge the device footprint, hindering miniaturization and high-density integration. In contrast, our flexible ISAC microdetector maintains a compact footprint without compromising responsivity. Moreover, statistical analysis of resistance and responsivity across multiple devices reveals a normal distribution, with mean values of $R$ = 230 kΩ and $R_V$ = 4 V/W, respectively (Supplementary Fig. 7). As a result, the combination of small device footprint, excellent performance, and high manufacturing yield highlights the potential of this PPAC design for scaling into large-scale, multi-pixel flexible THz detector arrays.

Frequency selectivity and polarization sensitivity are pivotal attributes of the flexible ISAC microdetector. To investigate the spectral response characteristics, we measured the device photocurrent across the 0.25–4.24 THz range and evaluated the corresponding responsivity. As illustrated in Fig. 3d, the measured responsivity exhibits a distinct frequency dependence (blue dots) that aligns well with the simulated THz absorption spectrum (red curve). This result confirms the frequency-resolved detection capability of the device, which is attributed to the frequency-dependent plasmon polariton resonance of the graphene rectangle PPACs. Furthermore, the anisotropic resonance of the rectangle PPACs endows the flexible ISAC microdetector with inherent polarization sensitivity. We employed a half-wave plate to modulate the polarization direction of the incident linearly polarized THz waves. The polarization angle is defined as 0° when the polarization aligns parallel to the long axis of the rectangle PPAC. As depicted in Fig. 3e, the responsivity of the device exhibits a dipole-like dependence on the polarization angle, forming an inverted figure-eight pattern (blue dots). This trend agrees excellently with the simulated polarization-resolved THz absorption intensity of the rectangle PPAC array (red curve). For comparison, we fabricated a flexible THz microdetector based on circular PPAC arrays and characterized its polarization response (Supplementary Fig. 8). As expected, both the experimental responsivity and the simulated absorption spectrum show negligible variation with polarization angle, indicating a lack of polarization sensitivity. These findings confirm that the polarization capability of the flexible ISAC microdetector stems from the anisotropic plasmonic resonance of the rectangle PPAC arrays. Under 2.52 THz illumination, the flexible ISAC microdetector exhibits a polarization ratio (PR) of 2.5. This value can be further enhanced by increasing the aspect ratio of the rectangle PPACs. Although the current PR is modest, to the best of our knowledge, this represents the first reported antenna- and metasurface-free, polarization-sensitive flexible THz microdetector based on 2D materials (Supplementary Table 1) [14–16, 26,

[27]. Collectively, these results highlight the geometric tunability of PPACs as a robust strategy for tailoring device performance to specific THz frequencies and polarization states, rendering it highly attractive for multi-band, tunable, and polarization-sensitive flexible THz detection.

Response speed is a critical parameter for the flexible ISAC microdetector. To obtain the response time of the device, we modulated the incident THz radiation using an optical chopper and recorded the photocurrent as a function of the modulation frequency. The response time is defined as the reciprocal of the modulation frequency at which the photocurrent falls to -3dB. As presented in Fig. 3f, the device exhibits a response time of approximately 62 ms, representing superior performance among state-of-the-art flexible THz detectors based on low-dimensional materials (Supplementary Table 1)[14–16, 26–33]. This rapid response arises from the high carrier mobility of graphene combined with the efficient PTE mechanism. Typically, graphene-based PTE detectors achieve response times below the microsecond range, with intrinsic speeds potentially reaching the nanosecond or picosecond regime[23, 34, 35]. The millisecond-scale response observed in our flexible ISAC microdetector is attributed to THz absorption by the PET substrate. Conventionally, graphene PTE detectors are fabricated on $Si/SiO_2$ substrates, which exhibit minimal THz absorption and thus exert negligible influence on response speed[26]. In contrast, THz absorption by the PET substrate induces uniform heating of carriers within the graphene channel, which enhances carrier scattering rates and impedes carrier diffusion toward the source and drain electrodes. Therefore, most flexible detectors reported to date operate with response times in the millisecond-to-second range (Supplementary Table 1)[14–16, 26–33]. Notably, our previous work demonstrated that optimizing device geometry, such as shortening the channel length and implementing asymmetric electrode designs, can enhance carrier mobility and further reduce response times[26].

Mechanical flexibility and stability are essential for the practical applications in flexible electronics. To evaluate these attributes, we characterized the mechanical stress and 2.52 THz response of the flexible ISAC microdetector under uniaxial bending strain. As depicted in Fig. 3g, the device withstands bending strains up to 3.6%—the mechanical limit of the bending machine—without exhibiting fracture or crack formation, demonstrating its exceptional flexibility. The THz response remains stable as the strain increases from 0% to 0.8%, ensuring reliable operation when conformally attached to the complex curvature of the human body. As the strain approaches 3.6%, the photocurrent experiences a moderate attenuation of 11.9%. This degradation is attributed to the strain-induced reduction in graphene conductivity[36] and the redshift of the PPAC plasmonic resonance peak (Fig. 1g). The latter gives rise to a reduction in the THz absorption of the graphene rectangle PPAC at 2.52 THz, thereby degrading the response of device. Interestingly, in applications of THz image sensing and classification based on a CNN, strain-induced photocurrent variation can in fact serve as an effective feature of image information, enhancing the discriminative ability of the system under varying strained conditions. To assess long-term mechanical stability, we performed a fatigue test repeatedly over 2000 bending times at 1.3% strain, recording the photocurrent every 200 cycles. The device maintains a highly stable output with the photocurrent decaying by only 1.2% after 2000 cycles. This negligible degradation is likely due to minor damage at the graphene-electrode contact interfaces. These results underscore the superior mechanical flexibility and stability of the flexible ISAC microdetector, confirming its significant potential for next-generation flexible wearable electronics.

**Sensing and classification of 2D THz ISAC smart wristband**

The sensing and identification capabilities of the flexible ISAC microdetector enable the classification of internal circuit faults, highlighting the potential of the 2D THz smart wristband for quality control and fault diagnosis in micro-electronics. To evaluated the classification capability of the device, THz responses from internal circuit faults were encoded into hybrid optical signals using the flexible ISAC microdetector and then used to train a CNN. As a proof of concept, a "CPU" pattern with conductive Au traces was fabricated on high-resistivity silicon substrate, and concealed behind an opaque high-resistivity silicon wafer to simulate an internal circuit inspection scenario (Fig. 4a). Artificial open- and short-circuit region were introduced into the design (bottom panel in Fig. 4a). Typically, such internal circuit faults are detected by 2D raster scanning of the "CPU" pattern using a THz dual-focus scanning imaging system (see Methods and Supplementary Fig. 9). By collecting the position-dependent photocurrent generated by THz waves transmitted through the silicon and the "CPU" pattern, a 2D image can be reconstructed to identify fault locations and types (upper left panel in Figs. 4b and 4c). However, this pixel-by-pixel scanning approach is time-consuming and inefficient for high-throughput semiconductor inspection.

In contrast, the flexible ISAC microdetector combined with a deep-learning-based CNN algorithm significantly enhances the detection efficiency for classifying open and short circuit faults[37]. We constructed an end-to-end CNN architecture to train on diverse THz signals acquired by the device under varying conditions, as illustrated in Fig. 4d. Specifically, THz signals carrying information about open-circuit (labeled "0")/ short-circuit (labeled "1") regions were detected using the flexible ISAC microdetector under different polarization angles (0° and 90°) and strain conditions (0% and 3.6%). For each condition, THz signals from 16 random locations in open-circuit/ short-circuit regions were detected and repeated 20 times, yielding a total dataset of 320 short-circuit and 320 open-circuit samples, each in the form of a 4×1 vector. Every sample signal was reshaped into a 2×2×1 feature vector as input to the CNN for circuit fault sensing and classification. The CNN consists of five convolutional layers for feature extraction, followed by a fully connected layer with dropout for feature fusion, and a Softmax classifier for final classification. To prevent feature information loss, pooling layers were omitted, and zero-padding was applied at the boundaries of each convolutional layer. The first layer utilizes a single filter to map the raw and simple input data into an 8×8×1 feature map, projecting it into a higher-dimensional spatial domain to uncover fundamental signal patterns. The second layer expands the number of filters to 16 to extract multiple primary feature combinations, such as edges and corners in different forms. The subsequent two convolutional layers further double the number of feature maps to 32 and 64, respectively, aiming to perform high-capacity and high-resolution feature mining. These layers employ numerous filters to capture subtle and complex discriminative patterns critical for accurately distinguishing fault types. The fifth layer reduces the filter count back to 32, compressing the rich feature set from the previous layer to extract the most robust and high-level features while controlling model complexity to mitigate overfitting. All convolutional kernels were set to 2×2 with a stride of 1. Each convolution operation was followed by a ReLU activation function and batch normalization to enhance nonlinear expression and improve training stability. Additionally, a 1×1 convolution layer with 32 filters was incorporated to fuse feature maps and reduce computational costs. Finally, the output feature maps were flattened into a vector and fed into a fully connected layer with 1024 neurons, which outputs the classification probabilities for open-circuit (labeled "0") and short-circuit (labeled "1") faults via a Softmax function.

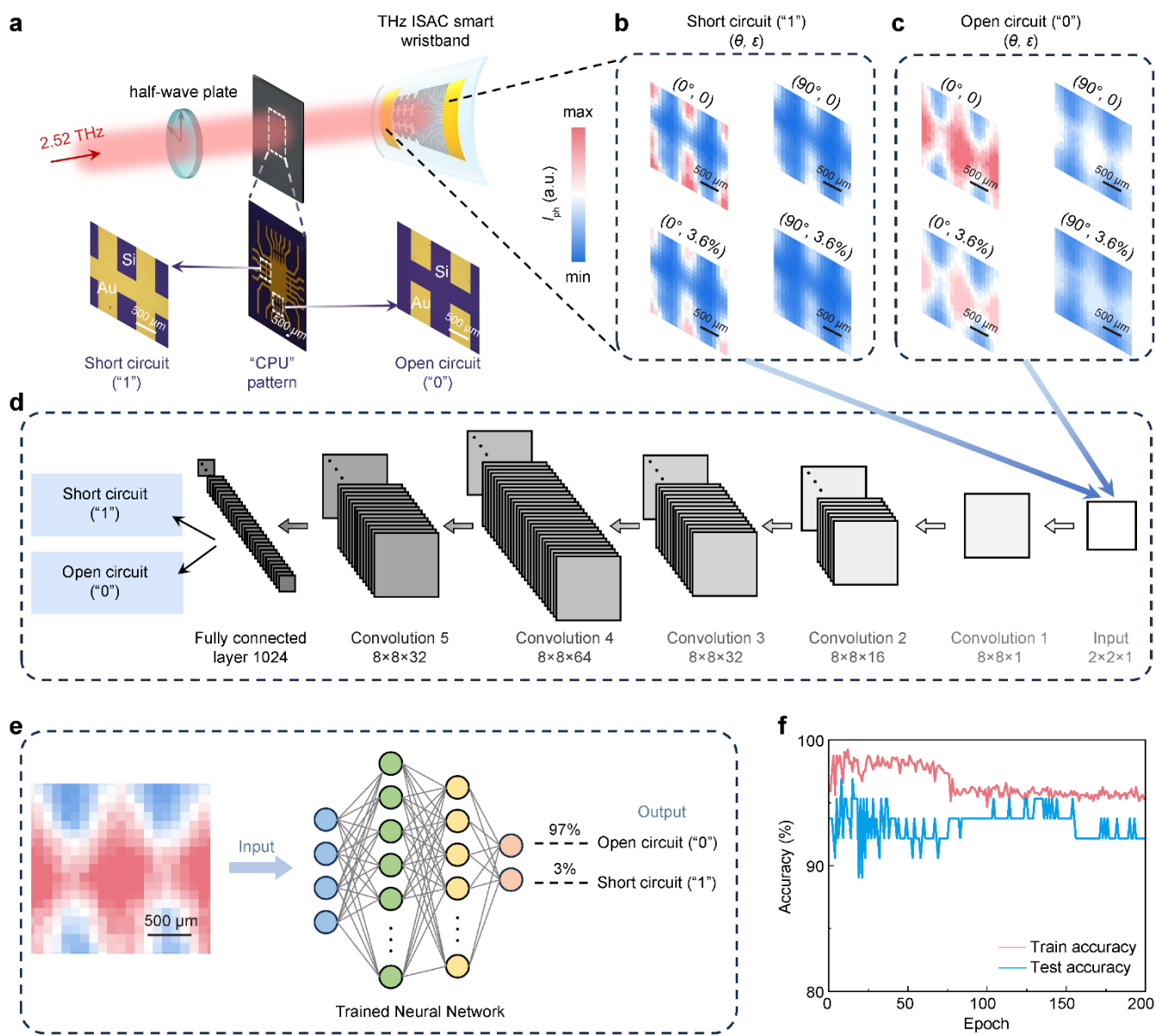

**Fig. 4 | Sensing and classification of the internal circuit faults by 2D THz ISAC wristband. a**, Schematic illustration of THz detection system for classifying internal circuit faults in chips. A half-wave plate is positioned at the emission port to adjust the polarization angle of the THz wave incident on the surface of the sample. The sample consists of a "CPU" pattern concealed behind an opaque silicon wafer (bottom middle panel), with artificially introduced short-circuit (bottom left panel) and open-circuit (bottom right panel) regions within the circuit. **b,c**, Schematic illustrating the THz signals carrying information about short-circuit and open-circuit regions are detected and encoded into 2×2×1 feature vector under varying polarization and strain conditions, which input to a CNN for training. The upper left panel in **b** and **c** depict photocurrent images of short-circuit and open-circuit regions using dual-focus scanning imaging system at $\theta = 0°$ and $\varepsilon = 0$, respectively. **d**, Detailed segment of the CNN architecture used for feature extraction and classification. **e**, Classification of circuit faults by a trained CNN. The corresponding fault type of the output neuron with the highest probability is the predicted result. **f**, Classification accuracy of CNN versus epoch.

The CNN was implemented using the TensorFlow 2.0 framework and trained on an NVIDIA 4090 GPU. Network weights were initialized using a truncated normal distribution, and the model was optimized using the Adam optimizer to minimize the cross-entropy loss function, with an initial learning rate of $10^{-4}$ and a dropout rate of 0.5. Using the trained CNN model, THz signals carrying information about unknown circuit fault region was classified, as present in Fig. 4e. We performed 10-fold cross-validation on the sample data, represented the test results as the average of the folds, and reported the final results as the mean over all test samples. The test results demonstrate that the trained CNN model can distinguish open-circuit and short-circuit faults with an accuracy of up to 97%, with a misclassification rate of only 3% (Fig. 4f). The training loss decreased rapidly within 25 epochs and converged by 50 epochs, indicating excellent classification performance and robustness (Supplementary Fig. 10). This approach significantly outperforms traditional point-scanning imaging strategies, greatly improving identification efficiency while maintain high accuracy, which offers a feasible technical route for high-throughput and non-destructive quality inspection and fault diagnosis in micro/nano-electronic circuits.

## Encrypted communication of 2D THz ISAC smart wristband

The flexible ISAC microdetector exhibits superior polarization sensitivity and distinct THz response characteristics, positioning it as a promising candidate for secure encrypted communication, which underscores the broad applicability of the 2D THz smart wristband in 6G networks and next-generation wearable electronics. Figure 5a illustrates the proposed dual-channel encrypted communication system, comprising THz signal modulation and signal reception. Within this system, Channel 1 and Channel 2 serve as independent transmission pathways for modulating and encoding the THz signals. Specifically, Channel 1 controls the on/off state of the THz wave to transmit the raw message, while Channel 2 modulates the polarization angle to carry the security key. By superimposing the raw message and the security key via the simultaneous manipulation of both channels, an encrypted THz signal is generated. The dual-channel current-amplitude decoding algorithm is implemented as follows (Fig. 5b): (i) a logic "0" is represented by Channel 1 in the off state with Channel 2 set at $\theta = 0°$ or $90°$; (ii) a logic "1" corresponds to Channel 1 in the on state with Channel 2 at $\theta = 90°$; and (iii) a logic "2" is defined by Channel 1 in the on state with Channel 2 at $\theta = 0°$. Since the dark current remains lower than the photocurrent, and the photocurrent at $\theta = 0°$ exceeds that at $\theta = 90°$, the logic states can be accurately distinguished based on photocurrent amplitude. At the signal receiver, the 2D THz ISAC smart wristband, worn on a simulated curved surface of the human wrist, converts the incoming THz signals into electrical signals in real time, which are subsequently processed and output to a display terminal. To demonstrate this dual-channel encrypted communication, the ASCII codes for "SYSU" and "NANO" were employed as the raw message and the security key, respectively, and encoded via Channel 1 and Channel 2. The two signals were superimposed to generate the encrypted signal transmission, as depicted in the "Input" and "Output" in Fig. 5c, which represent the encrypted optical signal and the corresponding photocurrent response, respectively. By applying the current-amplitude decoding algorithm (Fig. 5b), the raw message was successfully decoupled from the security key, enabling the accurate retrieval of the original "SYSU" information (Fig. 5c). These results not only establish the flexible ISAC microdetector as a vital component for future encrypted communication systems but also highlight its immense potential in driving the evolution of 6G security architectures and innovative flexible intelligent electronics.

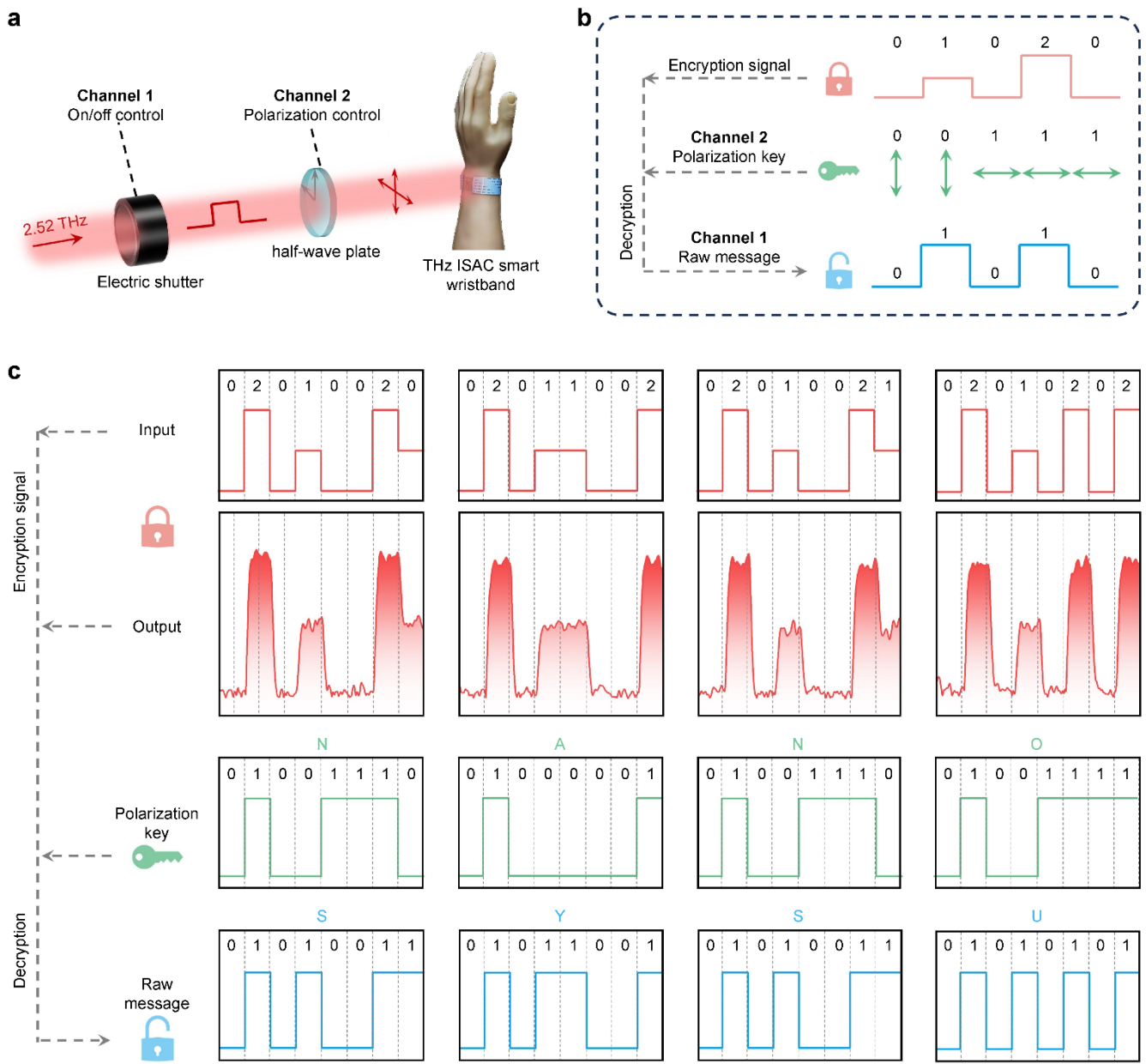

**Fig. 5 | Dual-channel encryption and decryption by the 2D THz ISAC wristband. a**, Schematic illustration of THz signal transmission system for dual-channel encrypted communication. **b**, Demonstration of current-amplitude decoding algorithm for encryption and decryption procedures of Channel 1 and Channel 2. **c**, Encrypted communication processes encoded for "SYSU". Encryption signal represents the photocurrent signal detected by the device after being modulated and encoded through Channel 1 and Channel 2. The ASCII code of the key file "NANO" is used for decrypting the raw message "SYSU" from the encryption signals.

## Conclusion

In summary, we have demonstrated a flexible ISAC microdetector based on graphene rectangle PPAC arrays, serving as the core component of a compact 2D THz ISAC smart wristband. Leveraging the plasmonic resonance modes within the PPACs, the device enables self-powered, polarization-sensitive and frequency-selective THz detection without the need for external antennas. Notably, to the best of our knowledge, it is the first reported polarization-sensitive 2D flexible THz detector that operates without integrated antennas or metasurfaces. The self-powered flexible device exhibits superior performance characteristics, including high responsivity up to 6.5 V/W, a fast response time of ~62 ms, and exceptional mechanical robustness maintaining stability over 2000 bending cycles. Impressively, the smart wristband achieves the monolithic integration of THz sensing and communication. For sensing, the device utilizes the polarization and strain response encoding of the detector, enabling a CNN to perform circuit fault sensing and classification with an accuracy of up to 97%. For communication, the device facilitates secure encrypted transmission and decryption through a dual-channel encoding scheme under simulated human body-wearing conditions. This work opens a promising avenue for the development of compact, high-efficiency flexible electronics tailored for the 6G IoT, semiconductor inspection and national defense. We anticipate that integrating THz ISAC functionalities into flexible smart wristbands, particularly when synergized with artificial intelligence and human-computer interaction technologies, will play a pivotal role in meeting the core demands of intelligent human-machine-object interconnectivity in the upcoming 6G era.

## Methods

### Graphene transfer and PPAC array fabrication

The PET substrate with a thickness of 180 μm, exhibits excellent stability, flexibility, and insulation properties as a flexible substrate. A monolayer graphene film with a polymethyl methacrylate (PMMA) coating layer (Jiangsu XFNANO Materials Tech. Co., Ltd.) was transferred onto the PET substrate via a wet transfer technique. The PMMA layer was then dissolved by immersing the sample in acetone for one hour. The monolayer graphene was patterned into rectangle PPAC array using a ultraviolet maskless lithography machine (TuoTuo Technology, UV Litho-ACA) and subsequently etched by oxygen plasma ($O_2$ at 400 Pa, power of 18 W, etching for 1 min). The photoresist was removed in acetone for one hour. The obtained PPAC arrays were characterized using optical microscopy, confocal Raman spectrometer, and THz-TDS.

### Characterizations

The morphology and Raman spectrum of graphene were characterized using a confocal Raman spectrometer (Renishaw, Invia Reflex). The absorption spectra of graphene rectangle PPACs were measured using a THz-TDS (BATOP, TDS-1008). The electrical transport characteristics of the flexible ISAC microdetector were tested using a source meter (Tektronix, Keithley 2636b). A far-infrared gas laser (Edinburgh Instruments, FIRL 100) and a microwave signal source (Rohde & Schwarz, SMB 100A) integrated with a frequency multiplier (VDI, SGX) and a tripler, with output frequencies of 4.24 THz, 3.11 THz, 2.52 THz, 1.84 THz, 0.53 THz, 0.35 THz–0.25 THz were employed for THz detection. The modulation frequency of the THz wave was controlled by an optical chopper (SCITEC, 310CD). A low-noise current pre-amplifier (FEMTO, DLPCA-200) was employed to amplify the signals from the flexible ISAC microdetector, and subsequently a lock-in amplifier (Stanford, SR830) was used for read out the amplified signals. A semiconductor parameter analyzer (PDA FS-Pro) was utilized to assess the noise current spectrum of the device. The uniaxial stress applied to the flexible ISAC microdetector was achieved through a single column tabletop testing systems (INSTRON, 5943).

Photoresponse measurements and imaging

In the photoresponse measurement, a laser beam with a spot size of approximately 1 mm, modulated by an optical chopper at 35 Hz, was uniformly illuminated on the device. Fine-tunning was carried out to optimize the photocurrent. The generated photocurrent was subsequently transformed into a voltage signal using a current pre-amplifier and recorded by a lock-in amplifier. For 2D raster imaging of a "CPU" pattern, the transmitted light from the short-circuit or open-circuit region, illuminated by a laser beam modulated at 35 Hz, was directed onto the device to generate photocurrent. The "CPU" pattern was scanned using a 2D electrically controlled displacement stage. The corresponding photocurrents at each position of the "CPU" pattern were measured by a current pre-amplifier and recorded by a lock-in amplifier. A 2D image of the "CPU" pattern was then generated by mapping the recorded photocurrent values to their respective spatial coordinates.

Simulations

The absorption spectra of graphene rectangle PPACs on PET substrate were simulated using FDTD. In the simulation, the dielectric function of the PET substrate was adopted from literature[38]. The optical conductivity of the graphene in the THz region can be defined as,

$$\sigma = \frac{je^2 E_f}{\pi \hbar^2 (\omega + j\tau^{-1})} \qquad (8)$$

where Fermi level $E_f$ and scattering rate ($\hbar\Gamma = \hbar\tau^{-1}/2$) were set to -0.3 eV and 0.003 eV, respectively. The PPACs were irradiated with a perpendicular incident plane wave of linear polarization. Due to absorption in PET substrates within the THz region, conventional methods based on transmission and reflection monitors incorporate this substrate loss. To accurately determine the intrinsic absorption of graphene PPACs ($A_{gr}$), a method based on the thermal loss principle should be employed[39]. Specifically, the electric field components ($E_x$, $E_y$) within the graphene PPACs region were first obtained from the simulation. Using the surface conductivity $\sigma$ of graphene, the current density $\bm{J}$ can be expressed as $\bm{J} = \sigma\bm{E}$[40]. The total power absorbed by graphene PPACs, $P_{absorbed}$, was obtained by integrating the power loss density over the graphene region, which can be defined as,

$$P_{\text{absorbed}} = \int_0^x \int_0^y \frac{1}{2} \text{Re}(\sigma |E_x|^2 + \sigma |E_y|^2) dx dy \tag{9}$$

The total power incident on the entire simulation domain was $P_{\text{incident}} = I_0 \times S$, where $I_0 = 1/(2 \times \eta_0)$ is the incident power density, $\eta_0$ is the wave impedance of a uniform plane wave in free space, and $S$ is the area of the array unit. Finally, the absorption rate of the graphene PPACs, $A_{\text{gr}}$, was obtained by $A_{\text{gr}} = P_{\text{absorbed}} / P_{\text{incident}}$.

## Acknowledgements

The authors acknowledge support from the National Key Basic Research Program of China (grant nos. 2024YFA1208500, 2024YFA1208501, and 2025YFA1213200), the National Natural Science Foundation of China (grant no. 92463308), Guangdong Basic and Applied Basic Research Foundation (grant no. 2023A1515011876), and China Postdoctoral Science Foundation (grant no. 2025M780806).